hep-lat/9509065   19 Sep 95

# Heavy-light meson decay constants from NRQCD: an analysis of the 1/M corrections.

Presented by S. Collins[a] *

[a]SCRI, Florida State University, Tallahassee, Fl 32306-4052, USA

We present *preliminary* results for the decay constants of heavy-light mesons using NRQCD heavy and tadpole improved Clover light quarks. A comparison is made with data obtained using Wilson light quarks. We present an analysis of the 1/M corrections to the decay constants in the static limit and compare with the predictions of HQET.

## 1. Introduction

We present an analysis of the mass dependence of the heavy-light (HL) meson decay constants. We use NRQCD for the heavy quark and include all 1/M corrections to the static limit. The corresponding NRQCD action is

$$\mathcal{L}_{NRQCD} = Q^{\dagger} \left\{ -D_t + \frac{D^2}{2M_Q^0} + \frac{g}{2M_Q^0}\sigma \cdot B \right\} Q. \quad (1)$$

In order to calculate matrix elements, and thus decay constants, to this order requires several operators. At tree level, these are obtained by relating the 4-component heavy quark field in full QCD ($q_h$) to the 2-component NRQCD field ($Q$) using the Foldy-Wouthuysen transformation (see [1]). Beyond tree level all operators with the same quantum numbers can mix under renormalisation. Thus to O($\alpha$/M) the basis of operators for the axial-pseudoscalar current, $\bar{q}\gamma_5\gamma_0 q_h$ is given by [2]

$$\mathcal{O}_1 = \bar{q}\gamma_5\gamma_0 Q \quad \mathcal{O}_2 = \bar{q}\gamma_5\gamma_0 \frac{\gamma_i D_i}{2M_Q^0}Q \quad (2)$$

$$\mathcal{O}_3 = \bar{q}\gamma_5\gamma_0 \frac{\overleftarrow{D_i}\,\gamma_i}{2M_Q^0}Q \quad (3)$$

Using translation invariance, currents involving $\mathcal{O}_2$ and $\mathcal{O}_3$ are identical on the lattice so it is only necessary to calculate matrix elements of the tree level operators $\mathcal{O}_1$ and $\mathcal{O}_2$. Similarly, the basis of operators for the vector current, $q\gamma_i q_h$,

$$\mathcal{O}_1 = \bar{q}\gamma_j Q \quad \mathcal{O}_2 = \bar{q}\gamma_j \frac{\gamma_i D_i}{2M_Q^0}Q \quad (4)$$

$$\mathcal{O}_3 = \bar{q}\frac{D_j}{2M_Q^0}Q \quad \mathcal{O}_4 = \bar{q}\frac{\overleftarrow{D_j}}{2M_Q^0}Q \quad (5)$$

$$\mathcal{O}_5 = \bar{q}\frac{\overleftarrow{D_i}\,\gamma_i}{2M_Q^0}\gamma_j Q \quad (6)$$

is reduced, for the numerical computation, to $\mathcal{O}_1$, $\mathcal{O}_2$ and $\mathcal{O}_3$, where $\gamma$ matrix properties have also been used.

In the same way as for the NRQCD action, the HL decay constants can be parameterised in the heavy quark limit in terms of an expansion in 1/M. To O(1/M)

$$f\sqrt{M} = (f\sqrt{M})^{\infty}\left(1 + \frac{c_P}{M} + O(\frac{1}{M^2})\right) \quad (7)$$

Note that (7) is valid for both the $PS$ and $V$ decay constants; we define $f\sqrt{M}_V \equiv \langle V|V|0\rangle/\sqrt{M_V}$. The coefficient $c_P$ is determined by nonperturbative contributions arising from the hyperfine interaction ($G_{hyp}$) and the kinetic energy of the heavy quark ($G_{kin}$), which appear in $\mathcal{L}_{NRQCD}$, and the corrections to the current ($G_{corr} \propto \langle \mathcal{O}_2|PS\rangle$) at tree level for $f_{PS}$:

$$c_P = G_{kin} + 3d_M G_{hyp} + d_M G_{corr}/6 \quad (8)$$

where $d_M = 3$ and $-1$ for $PS$ and $V$ mesons, respectively. From (8) the O(1/M) coefficient

---

for the spin-average of the decay constants, $(f\sqrt{M}_{\mathrm{PS}} + 3f\sqrt{M}_V)/4$ is

$$c'_P = G_{kin} \qquad (9)$$

Similarly, the ratio of decay constants, $f\sqrt{M}_{\mathrm{PS}}/f\sqrt{M}_V$, is 1 in the $M_Q^0 = \infty$ limit, and the O(1/M) coefficient $c''_P$ is given by

$$c''_P = 12G_{hyp} + 2G_{corr}/3. \qquad (10)$$

Thus, by computing various combinations of the PS and V currents the contributions from each O(1/M) term in the NRQCD action and matrix elements can be obtained. HQET provides an analogous decomposition of these coefficients (see [2] for a derivation of (8)). However, since HQET constructs an effective theory in terms of the heavy quark pole mass, as opposed to the bare quark mass in NRQCD, only physical combinations of $G_i$, ie $c_P$, $c''_P$ and $G_{kin}$, can be compared. Naively, these coefficients are expected to be $O(\Lambda_{\mathrm{QCD}}) \sim 200 - 500$ MeV and the corrections to the static limit at $M_B$ $O(\Lambda_{\mathrm{QCD}}/M_B) \sim 10\%$.

## 2. Computational Details

The simulation was performed on $100\ 16^3 \times 32$ lattices at $\beta = 5.6$ with two flavours of dynamical staggered fermions, generated by the HEMCGC collaboration [3]. The light quark propagators were generated using the Clover action with tadpole improvement, at three masses of light quark. For a comparison between using Clover and Wilson light quarks, two of the clover kappa values, $\kappa = 0.1385$ and $0.1401$ were matched to the Wilson kappa values $0.1585$ and $0.1600$ respectively using pion masses. Details of the Clover light spectroscopy results can be found in [4]. The heavy quark propagators were computed over a range of values of $M_Q^0$ between 0.8 and 10.0; the propagator in the static limit was also calculated. We will use a *nominal* value of $a^{-1} = 2.0$ GeV for this ensemble to present *preliminary* quantitative results.

We extracted the PS and V matrix element amplitudes in the standard way. We performed a simultaneous one exponential fit to SL and SS correlators, where the smearing function is the g.s. hydrogenic radial wavefunction with a radius of 3.0. Further details of our method and analysis can be found in [1]. The effective masses of the SL and SS pseudoscalar correlators for $M_Q^0 = 10.0$ and $\kappa_l = 0.1385$ are shown in figure 1; the mass obtained from the fit to the propagators is indicated. Note, even for a meson mass $> 3M_B$ the signal does not fall into noise until $t \sim 20$, compared to $t \sim 12$ for the static case.

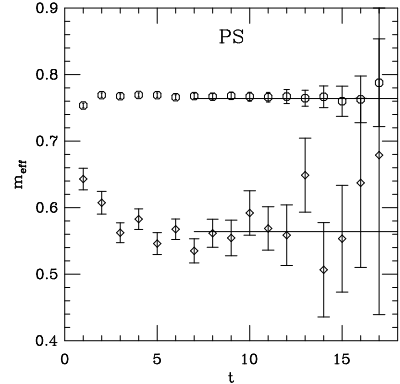

Figure 1. $m_{eff}$ for SL is offset by 0.2.

The data at different $M_Q^0$ and $\kappa_l$ are correlated and we found the optimal fitting range to be $7-20$ for PS and V correlators for $\kappa_l = 0.1385$ and $0.1393$ and $9-20$ for $\kappa_l = 0.1401$, for all $M_Q^0$. For the static case we used $8 - 12$. We computed the corrections to the currents separately by fitting to the ratio of the SL correlators with $\mathcal{O}_2$ and $\mathcal{O}_1$ at the sink, and in addition for the V meson the ratio of $\mathcal{O}_3$ and $\mathcal{O}_1$. The perturbative calculation of $Z_A$ and $Z_V$ is in progress, and has not yet been completed to include $\mathcal{O}_3$; we present the results for the tree-level operators.

## 3. Results

The PS decay amplitude with the tree level current correction included, $Z_A f\sqrt{M}_{\mathrm{PS}}^{corr}$, is plotted against the inverse meson mass ($M_2$) in figure 2 for $\kappa = 0.1385$. The result in the static limit is also shown. We performed a correlated fit to the data using the functional form

$$f\sqrt{M} = C_0 + \frac{C_1}{M} + \frac{C_2}{M^2} + \frac{C_3}{M^3}. \qquad (11)$$

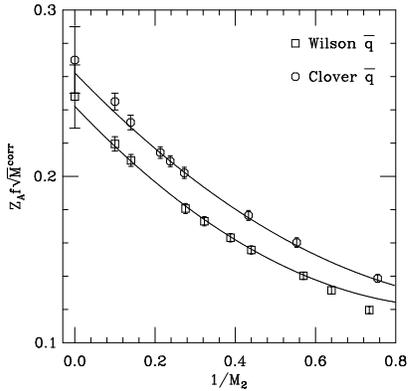

Figure 2.

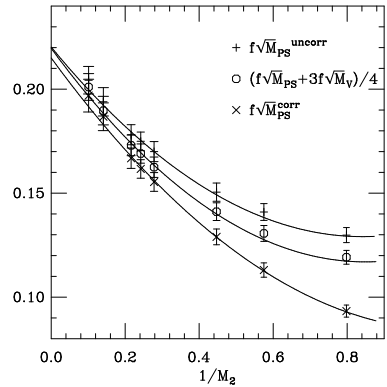

Figure 3.

| | range | Q | $C_0$ | $C_1$ | $C_2$ |
|---|---|---|---|---|---|
| C | 1-8 | 0.2 | 0.261(4) | -0.25(1) | 0.11(1) |
| W | 1-7 | 0.7 | 0.242(4) | -0.25(1) | 0.13(1) |

Table 1
$\kappa_l = 0.1385$

The fitting range was varied keeping the starting point fixed at the heaviest data point, $M_Q^0 = 10.0$. Beginning with a linear function, the fitting procedure was repeated adding quadratic and cubic terms. In order to have confidence in the value for $C_1$ a quadratic fit is required.

We found it was possible to perform a linear fit to $Z_A f \sqrt{M}_{PS}^{corr}$ for $M_Q^0 \gtrsim 3.0$. A quadratic function is necessary to fit to the rest of the data, ie there are significant O($1/M^2$) contributions to $Z_A f \sqrt{M}_{PS}^{corr}$ in the region of the B meson mass. Table 1 gives the values of $C_i$ obtained from the best quadratic fit. The fit is shown in figure 2. The extrapolation to the static limit is consistent with the static result, as expected; the static case is the infinite mass limit of NRQCD.

We compared these results with those obtained using Wilson light quarks at $\kappa_l = 0.1585$, also shown in figure 2. The coefficients obtained from the best quadratic fit to $Z_A f \sqrt{M}_{PS}^{corr}$ are given in table 1. There is a significant increase in $Z_A f \sqrt{M}_{PS} \mid_{M_Q^0 = \infty}$ of 10% using Clover light quarks compared to Wilson, and a similar de-crease in $C_1/C_0$.

Chirally extrapolating the Clover-NRQCD data and using the nominal value of $a^{-1}$, we find $f_B \sim 190(3)$ MeV and $f_{B_s}/f_B \sim 1.18(3)$ (errors are statistical). This can be very roughly compared with $f_B \sim 170$ MeV and $f_{B_s}/f_B \sim 1.3$ for Wilson light quarks, using the same $a^{-1}$.

A striking feature of the results is the large slope obtained; very roughly $C_1/C_0 \sim -1 a^{-1}$ or $\sim -2$ GeV, much larger than the naive expectation of $200 - 500$ MeV, and this leads to a large correction to the static limit at $M_B$ of $\sim 30$-40%. In addition, previous lattice calculations using Wilson or Clover fermions for the heavy quark around charm have found a significantly smaller slope, $C_1/C_0 \sim -1$GeV. To examine this further we obtained the individual contributions to the $PS$ matrix element amplitudes from each of the O($1/M$) interactions.

From eqn. (9) the linear part of the slope of $(f \sqrt{M}_{PS} + 3 f \sqrt{M}_V)/4$ arises purely from the kinetic energy of the heavy quark. Figure 3 presents the results for $\kappa_l = \kappa_c$; renormalisation factors are omitted and this introduces a O($\alpha_s$) $\sim 20\%$ error in the value of $C_1/C_0$ extracted. The $PS$ matrix element amplitude with and without the current correction, which include contributions from the other O($1/M$) terms, are also shown for comparison. It is clear from the large mass region where the slope is linear that $G_{kin}$ is much greater than

| | range | Q | $C_0$ | $C_1$ | $C_2$ |
|---|---|---|---|---|---|
| $f\sqrt{M}^{corr}_{PS}$ | 1-6 | 1.0 | 0.227(9) | -0.32(5) | 0.23(7) |
| $SA$ | 1-7 | 0.6 | 0.229(6) | -0.29(2) | 0.21(3) |
| $R$ | 1-8 | 1.0 | 0.98(1) | 0.33(10) | -0.20(10) |
| $R^{corr}$ | 1-8 | 1.0 | 1.00(2) | -0.18(9) | -0.19(9) |

Table 2

$\kappa_l = \kappa_c$

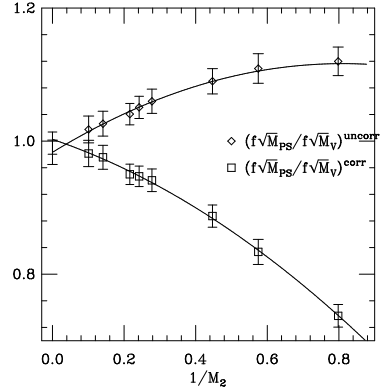

Figure 4. Errors in the $M_Q^0 = \infty$ extrapolation are indicated.

the contributions from $G_{hyp}$ or $G_{corr}$, and is the source of the large slope. The parameters obtained from the best fit to the data are shown in table (2) (SA). We find $G_{kin} = C_1/C_0 \sim -1.3a^{-1}$ or $\sim -2.6$ GeV.

To obtain the size of $G_{hyp}$ and $G_{corr}$ we consider the ratio of decay constants. $f\sqrt{M}_{PS}/f\sqrt{M}_V$ is presented in figure (4) both with and without the current correction. The data is consistent with 1 in both cases in the $M_Q^0 = \infty$ limit, as expected. From a quadratic fit to all the data points we find $c''_P \sim -0.2a^{-1}$ or $\sim -0.4$ GeV ($R^{corr}$ in table (2)). From eqn. (10) by omitting the current correction the linear coefficient is determined purely by $G_{hyp}$. We find $G_{hyp} \sim 0.03a^{-1}$ or $\sim 60$ MeV ($R$ in table (2)), and thus $G_{corr} \sim -0.8a^{-1}$ or $\sim -1.6$ GeV. Since we calculate the correction to the current separately, $G_{corr}$ can be more accurately extracted from fitting to $\langle O_2|PS\rangle/\langle O_1|PS\rangle$ directly. We find consistent results at this level of comparison, $G_{corr} \sim -0.6a^{-1}$.

A comparison can be made with the predictions of QCD sum rules. Using HQET Neubert [2] finds $c_P \sim -2.9(5)$ GeV, $c_{P''} \sim -0.9(1)$GeV, and $G_{kin} = -2.3(4)$ GeV, in good agreement with our results. As noted in section 1 $G_{hyp}$ and $G_{corr}$ differ. In fact in HQET $G_{corr} = -\bar{\Lambda}$, and in continuum NRQCD this becomes $G_{corr} = -E_{sim}$. Modulo lattice renormalisation factors we find agreement with this, $E_{sim} \sim 0.56$.

## 4. Conclusions

We presented an analysis of the mass dependence of HL decay constants using Clover light quarks and NRQCD heavy quarks to O(1/M). We find consistency between the extrapolation of the NRQCD data and the static case, and significant O(1/$M^2$) corrections to $f\sqrt{M}$ around $M_B$. The slope with $1/M_2$ is significantly larger than previous lattice calculations; this may be due to the use of too small a range of masses around the $D$ meson when using Wilson or Clover heavy quarks. In addition, we obtained the three O(1/M) contributions to the linear component of the slope separately and found $G_{kin}$ dominates. Good agreement is found between our results for $G_{kin}$ and physical combinations of $G_{hyp}$ and $G_{corr}$ and the predictions of HQET with QCD sum rules. The results indicate that the kinetic energy of the heavy quark gives rise to nonperturbative contributions to $f\sqrt{M}$ much greater than O($\Lambda_{QCD}$), contrary to naive expectations.

We found a significant increase in $f\sqrt{M}$ compared to using Wilson light quarks; it will be interesting to see if improved scaling behaviour is found using a lower $\beta$ value.

The authors would like to acknowledge the support under grants from NATO and DOE. The computations were performed on the CM-2 at SCRI.